\documentstyle[prd,aps,preprint,tighten,epsfig]{revtex}

\begin{document}

\draft

\title{Neutrino Telescopes as a Probe of Broken $\mu$-$\tau$ Symmetry}
\author{{\bf Zhi-zhong Xing}
\thanks{E-mail: xingzz@mail.ihep.ac.cn}}
\address{
CCAST (World Laboratory), P.O. Box 8730, Beijing 100080,
China \\
and
Institute of High Energy Physics, Chinese Academy of Sciences, \\
P.O. Box 918, Beijing 100049, China}

\maketitle

\begin{abstract}
It is known that neutrino oscillations may map $\phi^{}_e :
\phi^{}_\mu : \phi^{}_\tau = 1 : 2 : 0$, the initial flavor ratios
of ultrahigh-energy neutrino fluxes produced from a distant
astrophysical source, into $\phi^{\rm D}_e : \phi^{\rm D}_\mu :
\phi^{\rm D}_\tau = 1 : 1 : 1$ at the detector of a neutrino
telescope. We remark that this naive expectation is only valid in
the $\mu$-$\tau$ symmetry limit, in which two neutrino mixing
angles satisfy $\theta^{}_{13} = 0$ and $\theta^{}_{23} = \pi/4$.
Allowing for the slight breaking of $\mu$-$\tau$ symmetry, we find
$\phi^{\rm D}_e : \phi^{\rm D}_\mu : \phi^{\rm D}_\tau = (1 -2
\Delta) : (1 +\Delta) : (1 +\Delta)$ with $\Delta$ characterizing
the combined effect of $\theta^{}_{13} \neq 0$ and $\theta^{}_{23}
\neq \pi/4$. Current neutrino oscillation data indicate $-0.1
\lesssim \Delta \lesssim +0.1$. We also look at the possibility to
probe $\Delta$ by detecting the $\overline{\nu}^{}_e$ flux of
$E^{}_{\overline{\nu}^{}_e} \approx 6.3 ~ {\rm PeV}$ via the
Glashow resonance channel $\overline{\nu}^{}_e e \rightarrow W^-
\rightarrow ~ {\rm anything}$.
\end{abstract}


\newpage

\framebox{\Large\bf 1} ~ The solar \cite{SNO}, atmospheric
\cite{SK}, reactor \cite{KM} and accelerator \cite{K2K} neutrino
experiments have convinced us of the existence of neutrino
oscillations and opened a new window to physics beyond the
standard model. Given the basis in which the flavor eigenstates of
charged leptons are identified with their mass eigenstates, the
phenomenon of neutrino mixing can simply be described by a
$3\times 3$ unitary matrix $V$ which links the neutrino flavor
eigenstates $(\nu^{}_e, \nu^{}_\mu, \nu^{}_\tau)$ to the neutrino
mass eigenstates $(\nu^{}_1, \nu^{}_2, \nu^{}_3)$:
\begin{equation}
\left ( \matrix{\nu^{}_e \cr \nu^{}_\mu \cr \nu^{}_\tau \cr}
\right ) \; =\; \left ( \matrix{ V^{}_{e1} & V^{}_{e2} & V^{}_{e3}
\cr V^{}_{\mu 1} & V^{}_{\mu 2} & V^{}_{\mu 3} \cr V^{}_{\tau 1} &
V^{}_{\tau 2} & V^{}_{\tau 3} \cr} \right ) \left ( \matrix{
\nu^{}_1 \cr \nu^{}_2 \cr \nu^{}_3 \cr} \right ) \; .
\end{equation}
In the so-called standard parametrization of $V$ \cite{PDG},
$V^{}_{e2} = \sin\theta^{}_{12} \cos\theta^{}_{13}$, $V^{}_{e3} =
\sin\theta^{}_{13} e^{-i\delta}$ and $V^{}_{\mu 3} =
\sin\theta^{}_{23} \cos\theta^{}_{13}$. Here we have omitted the
Majorana CP-violating phases from $V$, because they are irrelevant
to the properties of neutrino oscillations to be discussed. A
global analysis of current experimental data (see, e.g., Ref.
\cite{Vissani}) points to $\theta^{}_{13} = 0$ and $\theta^{}_{23}
= \pi/4$, which motivate a number of authors to consider the
$\mu$-$\tau$ permutation symmetry for model building
\cite{Symmetry}. In the limit of $\mu$-$\tau$ symmetry, the
(effective) neutrino mass matrix $M^{}_\nu$ takes the form
\begin{equation}
M^{}_\nu \; = \; V^{}_0 \overline{M}^{}_\nu V^T_0 \; = \; \left (
\matrix{ a & b & b \cr b & c & d \cr b & d & c \cr} \right ) \; ,
\end{equation}
where $\overline{M}^{}_\nu \equiv {\rm Diag}\{ m^{}_1, m^{}_2,
m^{}_3\}$ with $m^{}_i$ (for $i=1,2,3$) being three neutrino
masses, and $V^{}_0$ is a special pattern of $V$ with
$\theta^{}_{13} = 0$ and $\theta^{}_{23} = \pi/4$. Note that
$\theta^{}_{12}$ is arbitrary and $\delta$ is not well-defined in
$V^{}_0$. The parameters $a$, $b$, $c$ and $d$ of $M^{}_\nu$ are
in general complex and can produce the desired Majorana
CP-violating phases for $V^{}_0$. Any slight breaking of
$\mu$-$\tau$ symmetry in $M^{}_\nu$ will result in non-vanishing
$\theta^{}_{13}$ and small departure of $\theta^{}_{23}$ from
$\pi/4$. Nontrivial $\delta$ can also be generated from the
breaking of $\mu$-$\tau$ symmetry, leading to the effect of CP
violation in neutrino oscillations.

The main purpose of this paper is to investigate how the effect of
$\mu$-$\tau$ symmetry breaking can show up at a neutrino
telescope. We anticipate that IceCube \cite{Ice} and other
second-generation neutrino telescopes \cite{Water} are able to
detect the fluxes of ultrahigh-energy $\nu^{}_e$
($\overline{\nu}^{}_e$), $\nu^{}_\mu$ ($\overline{\nu}^{}_\mu$)
and $\nu^{}_\tau$ ($\overline{\nu}^{}_\tau$) neutrinos generated
from very distant astrophysical sources. For most of the
currently-envisaged sources of ultrahigh-energy neutrinos
\cite{R}, a general and canonical expectation is that the initial
neutrino fluxes are produced via the decay of pions created from
$pp$ or $p\gamma$ collisions and their flavor content can be
expressed as
\begin{equation}
\left \{\phi^{}_e ~,~ \phi^{}_\mu ~,~ \phi^{}_\tau \right \} \; =
\; \left \{ \frac{1}{3} ~,~ \frac{2}{3} ~,~ 0 \right \} \phi^{}_0
\; ,
\end{equation}
where $\phi^{}_\alpha$ (for $\alpha = e, \mu, \tau$) denotes the
sum of $\nu^{}_\alpha$ and $\overline{\nu}^{}_\alpha$ fluxes, and
$\phi^{}_0 = \phi^{}_e + \phi^{}_\mu + \phi^{}_\tau$ is the total
flux of neutrinos and antineutrinos of all flavors. Due to
neutrino oscillations, the flavor composition of such cosmic
neutrino fluxes to be measured at the detector of a neutrino
telescope has been expected to be \cite{Pakvasa}
\begin{equation}
\left \{\phi^{\rm D}_e ,~ \phi^{\rm D}_\mu ,~ \phi^{\rm D}_\tau
\right \} \; = \; \left \{ \frac{1}{3} ~,~ \frac{1}{3} ~,~
\frac{1}{3} \right \} \phi^{}_0 \; .
\end{equation}
However, it is worth remarking that this naive expectation is only
true in the limit of $\mu$-$\tau$ symmetry (or equivalently,
$\theta^{}_{13} = 0$ and $\theta^{}_{23} = \pi/4$). Starting from
the hypothesis given in Eq. (3) and allowing for the slight
breaking of $\mu$-$\tau$ symmetry, we are going to show that
\begin{equation}
\left \{\phi^{\rm D}_e : ~ \phi^{\rm D}_\mu : ~ \phi^{\rm D}_\tau
\right \} \; = \; \left \{ \left (1 -2 \Delta \right ) : \left (1
+\Delta \right ) : \left (1 +\Delta \right ) \right \} \;
\end{equation}
holds to an excellent degree of accuracy, where $\Delta$
characterizes the effect of $\mu$-$\tau$ symmetry breaking (i.e.,
the combined effect of $\theta^{}_{13} \neq 0$ and $\theta^{}_{23}
\neq \pi/4$). We obtain $-0.1 \lesssim \Delta \lesssim +0.1$ from
current neutrino oscillation data. We find that it is also
possible to probe $\Delta$ by detecting the $\overline{\nu}^{}_e$
flux of $E^{}_{\overline{\nu}^{}_e} \approx 6.3 ~ {\rm PeV}$ via
the well-known Glashow resonance (GR) channel $\overline{\nu}^{}_e
e \rightarrow W^- \rightarrow ~ {\rm anything}$ \cite{Glashow} at
a neutrino telescope.

\vspace{0.5cm}

\framebox{\Large\bf 2} ~ Let us define $\phi^{(\rm D)}_\alpha
\equiv \phi^{(\rm D)}_{\nu^{}_\alpha} + \phi^{(\rm
D)}_{\overline{\nu}^{}_\alpha}$ (for $\alpha = e, \mu, \tau$)
throughout this paper, where $\phi^{(\rm D)}_{\nu^{}_\alpha}$ and
$\phi^{(\rm D)}_{\overline{\nu}^{}_\alpha}$ denote the
$\nu^{}_\alpha$ and $\overline{\nu}^{}_\alpha$ fluxes,
respectively. As for the ultrahigh-energy neutrino fluxes produced
from the pion-muon decay chain with $\phi^{}_{\nu^{}_\tau} =
\phi^{}_{\overline{\nu}^{}_\tau} = 0$, the relationship between
$\phi^{}_{\nu^{}_\alpha}$ (or
$\phi^{}_{\overline{\nu}^{}_\alpha}$) and $\phi^{\rm
D}_{\nu^{}_\alpha}$ (or $\phi^{\rm D}_{\overline{\nu}^{}_\alpha}$)
is given by $\phi^{\rm D}_{\nu^{}_\alpha} = \phi^{}_{\nu^{}_e}
P^{}_{e\alpha} + \phi^{}_{\nu^{}_\mu} P^{}_{\mu\alpha}$ or
$\phi^{\rm D}_{\overline{\nu}^{}_\alpha} =
\phi^{}_{\overline{\nu}^{}_e} \bar{P}^{}_{e\alpha} +
\phi^{}_{\overline{\nu}^{}_\mu} \bar{P}^{}_{\mu\alpha}$, in which
$P^{}_{\beta\alpha}$ and $\bar{P}^{}_{\beta\alpha}$ (for $\alpha =
e, \mu, \tau$ and $\beta = e$ or $\mu$) stand respectively for the
oscillation probabilities $P (\nu^{}_\beta \rightarrow
\nu^{}_\alpha)$ and $P (\overline{\nu}^{}_\beta \rightarrow
\overline{\nu}^{}_\alpha)$. Because the Galactic distances far
exceed the observed neutrino oscillation lengths,
$P^{}_{\beta\alpha}$ and $\bar{P}^{}_{\beta\alpha}$ are actually
averaged over many oscillations. Then we obtain
$\bar{P}^{}_{\beta\alpha} = P^{}_{\beta\alpha}$ and
\begin{equation}
P^{}_{\beta\alpha} \; = \; \sum^3_{i=1} |V^{}_{\alpha i}|^2
|V^{}_{\beta i}|^2 \; ,
\end{equation}
where $V^{}_{\alpha i}$ and $V^{}_{\beta i}$ (for $\alpha, \beta =
e, \mu, \tau$ and $i = 1, 2, 3$) denote the matrix elements of $V$
defined in Eq. (1). The relationship between $\phi^{}_\alpha$ and
$\phi^{\rm D}_\alpha$ turns out to be
\begin{equation}
\phi^{\rm D}_\alpha \; = \; \phi^{}_e P^{}_{e\alpha} + \phi^{}_\mu
P^{}_{\mu\alpha} \; .
\end{equation}
To be explicit, we have
\begin{eqnarray}
\phi^{\rm D}_e & = & \frac{\phi^{}_0}{3} \left (P^{}_{ee} + 2
P^{}_{\mu e} \right ) \; , \nonumber \\
\phi^{\rm D}_\mu & = & \frac{\phi^{}_0}{3} \left (P^{}_{e\mu} + 2
P^{}_{\mu \mu} \right ) \; , \nonumber \\
\phi^{\rm D}_\tau & = & \frac{\phi^{}_0}{3} \left (P^{}_{e\tau} +
2P^{}_{\mu \tau} \right ) \; .
\end{eqnarray}
It is then possible to evaluate the relative magnitudes of
$\phi^{\rm D}_e$, $\phi^{\rm D}_\mu$ and $\phi^{\rm D}_\tau$ by
using Eqs. (1), (6) and (8).

In order to clearly show the effect of $\mu$-$\tau$ symmetry
breaking on the neutrino fluxes to be detected at neutrino
telescopes, we define a small quantity
\begin{equation}
\varepsilon \; \equiv \; \theta^{}_{23} - \frac{\pi}{4} \; ,
~~~~~~~ (|\varepsilon| \ll 1) \; .
\end{equation}
Namely, $\varepsilon$ measures the slight departure of
$\theta^{}_{23}$ from $\pi/4$. Using small $\theta^{}_{13}$ and
$\varepsilon$, we may express $|V^{}_{\alpha i}|^2$ (for $\alpha
=e, \mu, \tau$ and $i=1,2,3$) as follows:
\begin{eqnarray}
\left [ \matrix{ |V^{}_{e1}|^2 & |V^{}_{e2}|^2 & |V^{}_{e3}|^2 \cr
|V^{}_{\mu 1}|^2 & |V^{}_{\mu 2}|^2 & |V^{}_{\mu 3}|^2 \cr
|V^{}_{\tau 1}|^2 & |V^{}_{\tau 2}|^2 & |V^{}_{\tau 3}|^2 \cr}
\right ] & = & \frac{1}{2} \left [ \matrix{ 2\cos^2\theta^{}_{12}
& 2\sin^2\theta^{}_{12} & 0 \cr \sin^2\theta^{}_{12} &
\cos^2\theta^{}_{12} & 1 \cr \sin^2\theta^{}_{12} &
\cos^2\theta^{}_{12} & 1 \cr} \right ] + \varepsilon \left [
\matrix{ 0 & 0 & 0 \cr -\sin^2\theta^{}_{12} &
-\cos^2\theta^{}_{12} & 1 \cr \sin^2\theta^{}_{12} &
\cos^2\theta^{}_{12} & -1 \cr} \right ]
\nonumber \\
& & + ~ \frac{\theta^{}_{13}}{2} \sin 2\theta^{}_{12} \cos\delta
\left [ \matrix{ 0 & 0 & 0 \cr 1 & -1 & 0 \cr -1 & 1 & 0 \cr}
\right ] + {\cal O}(\varepsilon^2) + {\cal O}(\theta^2_{13}) \; .
\end{eqnarray}
Combining Eqs. (6) and (10) allows us to calculate
$P^{}_{\beta\alpha}$. After a straightforward calculation, we
obtain
\begin{eqnarray}
P^{}_{ee} + 2 P^{}_{\mu e} & = & 1 - \varepsilon \sin^2
2\theta^{}_{12} + \frac{\theta^{}_{13}}{2} \sin 4\theta^{}_{12}
\cos\delta + {\cal O}(\varepsilon^2) + {\cal O}(\theta^2_{13}) \;
,
\nonumber \\
P^{}_{e\mu} + 2 P^{}_{\mu \mu} & = & 1 + \frac{\varepsilon}{2}
\sin^2 2\theta^{}_{12} - \frac{\theta^{}_{13}}{4} \sin
4\theta^{}_{12} \cos\delta + {\cal O}(\varepsilon^2) + {\cal
O}(\theta^2_{13}) \; ,
\nonumber \\
P^{}_{e\tau} + 2 P^{}_{\mu \tau} & = & 1 + \frac{\varepsilon}{2}
\sin^2 2\theta^{}_{12} - \frac{\theta^{}_{13}}{4} \sin
4\theta^{}_{12} \cos\delta + {\cal O}(\varepsilon^2) + {\cal
O}(\theta^2_{13}) \; .
\end{eqnarray}
Substituting Eq. (11) into Eq. (8), we arrive at
\begin{eqnarray}
\phi^{\rm D}_e & = & \frac{\phi^{}_0}{3} \left ( 1 - 2\Delta \right )
\; , \nonumber \\
\phi^{\rm D}_\mu & = & \frac{\phi^{}_0}{3} \left ( 1 + \Delta \right )
\; , \nonumber \\
\phi^{\rm D}_\tau & = & \frac{\phi^{}_0}{3} \left ( 1 + \Delta
\right ) \; ,
\end{eqnarray}
where
\begin{equation}
\Delta \; = \; \frac{1}{4} \left ( 2\varepsilon \sin^2
2\theta^{}_{12} - \theta^{}_{13} \sin 4\theta^{}_{12} \cos\delta
\right ) + {\cal O}(\varepsilon^2) + {\cal O}(\theta^2_{13}) \; .
\end{equation}
Eq. (5) is therefore proved by Eq. (12). One can see that
$\phi^{\rm D}_e + \phi^{\rm D}_\mu + \phi^{\rm D}_\tau =
\phi^{}_0$ holds. Some discussions are in order.

(1) The small parameter $\Delta$ characterizes the overall effect
of $\mu$-$\tau$ symmetry breaking. Allowing $\delta$ to vary
between $0$ and $\pi$, we may easily obtain the lower and upper
bounds of $\Delta$ for given values of $\theta^{}_{12}$ ($<
\pi/4$), $\theta^{}_{13}$ and $\varepsilon$: $-\Delta^{}_{\rm
bound} \leq \Delta \leq +\Delta^{}_{\rm bound}$, where
\begin{equation}
\Delta^{}_{\rm bound} = \frac{1}{4} \left ( 2|\varepsilon| \sin^2
2\theta^{}_{12} + \theta^{}_{13} \sin 4\theta^{}_{12} \right ) +
{\cal O}(\varepsilon^2) + {\cal O}(\theta^2_{13}) \; .
\end{equation}
It is obvious that $\Delta = -\Delta^{}_{\rm bound}$ when
$\varepsilon <0$ and $\delta =0$, and $\Delta = +\Delta^{}_{\rm
bound}$ when $\varepsilon >0$ and $\delta =\pi$. A global analysis
of current neutrino oscillation data \cite{Vissani} indicates
$30^\circ < \theta^{}_{12} < 38^\circ$, $\theta^{}_{13} <
10^\circ$ ($\approx 0.17$) and $|\varepsilon| < 9^\circ$ ($\approx
0.16$) at the $99\%$ confidence level, but the CP-violating phase
$\delta$ is entirely unrestricted. Using these constraints, we
analyze the allowed range of $\Delta$ and its dependence on
$\delta$. The maximal value of $\Delta^{}_{\rm bound}$ (i.e.,
$\Delta^{}_{\rm bound} \approx 0.098$) appears when
$|\varepsilon|$ and $\theta^{}_{13}$ approach their respective
upper limits and $\theta^{}_{12} \approx 33^\circ$ holds, as one
can clearly see from Fig. 1(A). Indeed, we find that
$\Delta^{}_{\rm bound}$ is not very sensitive to the variation of
$\theta^{}_{12}$ in its allowed region.

Provided $\theta^{}_{13} =0$ holds, we easily obtain
$\Delta^{}_{\rm bound} = 0.5|\varepsilon|\sin^2 2\theta^{}_{12}
<0.074$ when $\theta^{}_{12}$ approaches its upper limit. If
$\varepsilon =0$ (i.e., $\theta^{}_{23} =\pi/4$) holds,
nevertheless, we find $\Delta^{}_{\rm bound} = 0.25 \theta^{}_{13}
\sin 4\theta^{}_{12} <0.038$ as $\theta^{}_{12}$ approaches its
lower limit. We observe that $\Delta^{}_{\rm bound}$ is more
sensitive to the deviation of $\theta^{}_{23}$ from $\pi/4$.

(2) Of course, $\Delta =0$ exactly holds when $\theta^{}_{13} =
\varepsilon =0$ is taken. Because the sign of $\varepsilon$ and
the range of $\delta$ are both unknown, we are now unable to rule
out the nontrivial possibility $\Delta \approx 0$ in the presence
of $\theta^{}_{13} \neq 0$ and $\varepsilon \neq 0$. In other
words, $\Delta$ may be vanishing or extremely small if its two
leading terms cancel each other. It is straightforward to arrive
at $\Delta \approx 0$ from Eq. (13), if the condition
\begin{equation}
\frac{\varepsilon}{\theta^{}_{13}} \; =\; \cot 2\theta^{}_{12}
\cos\delta \;
\end{equation}
is satisfied. Due to $|\cos\delta| \leq 1$, Eq. (15) imposes a
strong constraint on the magnitude of
$\varepsilon/\theta^{}_{13}$. The dependence of
$\varepsilon/\theta^{}_{13}$ on $\delta$ is illustrated in Fig.
1(B), where $\theta^{}_{12}$ varies in its allowed range. One can
see that $|\varepsilon|/\theta^{}_{13} < 0.6$ is necessary to
hold, such that a large cancellation between two leading terms of
$\Delta$ is possible to take place. It should be remarked again
that the above result is a natural consequence of the assumption
made in Eq. (3) for the initial flavor ratios of ultrahigh-energy
neutrino fluxes.

The implication of Fig. 1 on high-energy neutrino telescopes is
two-fold. On the one hand, an observable signal of $\Delta \neq 0$
at a neutrino telescope implies the existence of significant
$\mu$-$\tau$ symmetry breaking. If a signal of $\Delta \neq 0$
does not show up at a neutrino telescope, on the other hand, one
cannot conclude that the $\mu$-$\tau$ symmetry is an exact or
almost exact symmetry. It is therefore meaningful to consider the
complementarity between neutrino telescopes and terrestrial
neutrino oscillation experiments \cite{Winter}, in order to
finally pin down the parameters of neutrino mixing and leptonic CP
violation.

(3) To illustrate, we define three neutrino flux ratios
\begin{eqnarray}
R^{}_e & \equiv & \frac{\phi^{\rm D}_e}{\phi^{\rm D}_\mu +
\phi^{\rm D}_\tau} \; ,
\nonumber \\
R^{}_\mu & \equiv & \frac{\phi^{\rm D}_\mu}{\phi^{\rm D}_\tau +
\phi^{\rm D}_e} \; ,
\nonumber \\
R^{}_\tau & \equiv & \frac{\phi^{\rm D}_\tau}{\phi^{\rm D}_e +
\phi^{\rm D}_\mu} \; ,
\end{eqnarray}
which may serve as the {\it working} observables at neutrino
telescopes \cite{XZ}. At least, $R^{}_\mu$ can be extracted from
the ratio of muon tracks to showers at IceCube \cite{Ice}, even if
those electron and tau events cannot be disentangled. Taking
account of Eq. (12), we approximately obtain
\begin{eqnarray}
R^{}_e & \approx & \frac{1}{2} ~ - ~ \frac{3}{2} \Delta \; ,
\nonumber \\
R^{}_\mu & \approx & \frac{1}{2} ~ + ~ \frac{3}{4} \Delta \; ,
\nonumber \\
R^{}_\tau & \approx & \frac{1}{2} ~ + ~ \frac{3}{4} \Delta \; .
\end{eqnarray}
It turns out that $R^{}_e$ is most sensitive to the effect of
$\mu$-$\tau$ symmetry breaking.

As a straightforward consequence of $\phi^{\rm D}_\mu = \phi^{\rm
D}_\tau$ shown in Eq. (12), $R^{}_\mu = R^{}_\tau$ holds no matter
whether $\Delta$ vanishes or not. This interesting observation
implies that the ``$\mu$-$\tau$" symmetry between $R^{}_\mu$ and
$R^{}_\tau$ is actually insensitive to the breaking of
$\mu$-$\tau$ symmetry in the neutrino mass matrix $M^{}_\nu$. If
both $R^{}_e$ and $R^{}_\mu$ are measured at a neutrino telescope,
one can then extract the information about $\Delta$ from the
difference of these two observables:
\begin{equation}
R^{}_\mu - R^{}_e \; =\; \frac{9}{4} \Delta \; .
\end{equation}
Taking $\Delta = \Delta^{}_{\rm bound} \approx 0.1$, we get
$R^{}_\mu - R^{}_e \lesssim 0.22$.

\vspace{0.5cm}

\framebox{\Large\bf 3} ~ We proceed to discuss the possibility to
probe the breaking of $\mu$-$\tau$ symmetry by detecting the
$\overline{\nu}^{}_e$ flux from distant astrophysical sources
through the Glashow resonance (GR) channel $\overline{\nu}^{}_e e
\rightarrow W^- \rightarrow ~ {\rm anything}$ \cite{Glashow}. The
latter can take place over a narrow energy interval around the
$\overline{\nu}^{}_e$ energy $E^{\rm GR}_{\overline{\nu}^{}_e}
\approx M^2_W/2m^{}_e \approx 6.3 ~ {\rm PeV}$. A neutrino
telescope may measure both the GR-mediated $\overline{\nu}^{}_e$
events ($N^{\rm GR}_{\overline{\nu}^{}_e}$) and the $\nu^{}_\mu +
\overline{\nu}^{}_\mu$ events of charged-current (CC) interactions
($N^{\rm CC}_{\nu^{}_\mu + \overline{\nu}^{}_\mu}$) in the
vicinity of $E^{\rm GR}_{\overline{\nu}^{}_e}$. Their ratio,
defined as $R^{}_{\rm RG} \equiv N^{\rm
GR}_{\overline{\nu}^{}_e}/N^{\rm CC}_{\nu^{}_\mu +
\overline{\nu}^{}_\mu}$, can be related to the ratio of
$\overline{\nu}^{}_e$'s to $\nu^{}_\mu$'s and
$\overline{\nu}^{}_\mu$'s entering the detector,
\begin{equation}
R^{}_0 \; \equiv \; \frac{\phi^{\rm
D}_{\overline{\nu}^{}_e}}{\phi^{\rm D}_{\nu^{}_\mu} + \phi^{\rm
D}_{\overline{\nu}^{}_\mu}} \; .
\end{equation}
Note that $\phi^{\rm D}_{\overline{\nu}^{}_e}$, $\phi^{\rm
D}_{\nu^{}_\mu}$ and $\phi^{\rm D}_{\overline{\nu}^{}_\mu}$ stand
respectively for the fluxes of $\overline{\nu}^{}_e$'s,
$\nu^{}_\mu$'s and $\overline{\nu}^{}_\mu$'s before the RG and CC
interactions occur at the detector. In Ref. \cite{BG}, $R^{}_{\rm
GR} = a R^{}_0$ with $a \approx 30.5$ has been obtained by
considering the muon events with contained vertices \cite{Beacom}
in a water- or ice-based detector. An accurate calculation of $a$
is certainly crucial for a specific neutrino telescope to detect
the GR reaction rate, but it is beyond the scope of this work.
Instead, here we concentrate on the possible effect of
$\mu$-$\tau$ symmetry breaking on $R^{}_0$.

Provided the initial neutrino fluxes are produced via the decay of
$\pi^+$'s and $\pi^-$'s created from high-energy $pp$ collisions,
their flavor composition can be expressed in a more detailed way
as follows:
\begin{equation}
\left \{\phi^{}_{\nu^{}_e} ,~ \phi^{}_{\overline{\nu}^{}_e} ,~
\phi^{}_{\nu^{}_\mu} ,~ \phi^{}_{\overline{\nu}^{}_\mu} ,~
\phi^{}_{\nu^{}_\tau} ,~ \phi^{}_{\overline{\nu}^{}_\tau} \right
\} \; = \; \left \{ \frac{1}{6} ~,~ \frac{1}{6} ~,~ \frac{1}{3}
~,~ \frac{1}{3} ~,~ 0 ~,~ 0 \right \} \phi^{}_0 \; .
\end{equation}
In comparison, the flavor content of ultrahigh-energy neutrino
fluxes produced from $p\gamma$ collisions reads
\footnote{Note that the dominant reaction to generate electron and
muon neutrinos in $p\gamma$ collisions is $p\gamma \rightarrow
\Delta^+ \rightarrow \pi^+ n$ with $\pi^+ \rightarrow \mu^+
\nu^{}_\mu$ and $\mu^+ \rightarrow e^+ \nu^{}_e
\overline{\nu}^{}_\mu$. There is no production of
$\overline{\nu}^{}_e$, because the produced neutrons can escape
the source before decaying \cite{Weiler}. In contrast, the numbers
of $\nu^{}_e$'s and $\overline{\nu}^{}_e$'s in Eq. (20) are
identical as a result of the equal amount of $\pi^+$'s and
$\pi^-$'s produced from inelastic $pp$ collisions.}
\begin{equation}
\left \{\phi^{}_{\nu^{}_e} ,~ \phi^{}_{\overline{\nu}^{}_e} ,~
\phi^{}_{\nu^{}_\mu} ,~ \phi^{}_{\overline{\nu}^{}_\mu} ,~
\phi^{}_{\nu^{}_\tau} ,~ \phi^{}_{\overline{\nu}^{}_\tau} \right
\} \; = \; \left \{ \frac{1}{3} ~,~ 0 ~,~ \frac{1}{3} ~,~
\frac{1}{3} ~,~ 0 ~,~ 0 \right \} \phi^{}_0 \; .
\end{equation}
For either Eq. (20) or Eq. (21), the sum of
$\phi^{}_{\nu^{}_\alpha}$ and $\phi^{}_{\overline{\nu}^{}_\alpha}$
(for $\alpha = e, \mu, \tau$) is consistent with $\phi^{}_\alpha$
in Eq. (3).

Due to neutrino oscillations, the $\overline{\nu}^{}_e$ flux at
the detector of a neutrino telescope is given by $\phi^{\rm
D}_{\overline{\nu}^{}_e} = \phi^{}_{\overline{\nu}^{}_e}
\bar{P}^{}_{ee} + \phi^{}_{\overline{\nu}^{}_\mu} \bar{P}^{}_{\mu
e}$. With the help of Eqs. (6), (10), (20) and (21), we explicitly
obtain
\begin{eqnarray}
\phi^{\rm D}_{\overline{\nu}^{}_e} (pp) ~ & = & ~
\frac{\phi^{}_0}{6} \left (1 - 2 \Delta \right ) \; ,
\nonumber \\
\phi^{\rm D}_{\overline{\nu}^{}_e} (p\gamma) ~ & = & ~
\frac{\phi^{}_0}{12} \left (\sin^2 2\theta^{}_{12} - 4 \Delta
\right ) \; .
\end{eqnarray}
The sum of $\phi^{\rm D}_{\nu^{}_\mu}$ and $\phi^{\rm
D}_{\overline{\nu}^{}_\mu}$, which is defined as $\phi^{\rm
D}_\mu$, has been given in Eq. (12). It is then straightforward to
calculate $R^{}_0$ by using Eq. (19) for two different
astrophysical sources:
\begin{eqnarray}
R^{}_0(pp) ~ & \approx & ~ \frac{1}{2} ~ - ~ \frac{3}{2} \Delta \;
,
\nonumber \\
R^{}_0(p\gamma) ~ & \approx & ~ \frac{\sin^2 2\theta^{}_{12}}{4} ~
- ~ \frac{4 + \sin^2 2\theta^{}_{12}}{4} \Delta \; .
\end{eqnarray}
This result indicates that the dependence of $R^{}_0(pp)$ on
$\theta^{}_{12}$ is hidden in $\Delta$ and suppressed by the
smallness of $\theta^{}_{13}$ and $\varepsilon$. In addition, the
deviation of $R^{}_0(pp)$ from $1/2$ can be as large as
$1.5\Delta^{}_{\rm bound} \approx 0.15$. It is obvious that the
ratio $R^{}_0(p\gamma)$ is very sensitive to the value of $\sin^2
2\theta^{}_{12}$. A measurement of $R^{}_0 (p\gamma)$ at IceCube
and other second-generation neutrino telescopes may therefore
probe the solar neutrino mixing angle $\theta^{}_{12}$ \cite{BG}.
Indeed, the dominant production mechanism for ultrahigh-energy
neutrinos at Active Galactic Nuclei (AGNs) and Gamma Ray Bursts
(GRBs) is expected to be the $p\gamma$ process in a tenuous or
radiation-dominated environment \cite{Berezinsky}. If this
expectation is true, the observation of $R^{}_0(p\gamma)$ may also
provide us with useful information on the breaking of $\mu$-$\tau$
symmetry.

\vspace{0.5cm}

\framebox{\Large\bf 4} ~ In summary, we have discussed why and how
the second-generation neutrino telescopes can serve as a striking
probe of broken $\mu$-$\tau$ symmetry. Based on the conventional
mechanism for ultrahigh-energy neutrino production at a distant
astrophysical source and the standard picture of neutrino
oscillations, we have shown that the flavor composition of cosmic
neutrino fluxes at a terrestrial detector may deviate from the
naive expectation $\phi^{\rm D}_e : \phi^{\rm D}_\mu : \phi^{\rm
D}_\tau = 1 : 1 : 1$. Instead, $\phi^{\rm D}_e : \phi^{\rm D}_\mu
: \phi^{\rm D}_\tau = (1 -2 \Delta) : (1 +\Delta) : (1 +\Delta)$
holds, where $\Delta$ characterizes the effect of $\mu$-$\tau$
symmetry breaking. The latter is actually a reflection of
$\theta^{}_{13} \neq 0$ and $\theta^{}_{23} \neq \pi/4$ in the
$3\times 3$ neutrino mixing matrix. We have examined the
sensitivity of $\Delta$ to the deviation of $\theta^{}_{13}$ from
zero and to the departure of $\theta^{}_{23}$ from $\pi/4$, and
obtained $-0.1 \lesssim \Delta \lesssim +0.1$ from current
neutrino oscillation data. We find that it is also possible to
probe the breaking of $\mu$-$\tau$ symmetry by detecting the
$\overline{\nu}^{}_e$ flux of $E^{}_{\overline{\nu}^{}_e} \approx
6.3 ~ {\rm PeV}$ via the Glashow resonance channel
$\overline{\nu}^{}_e e \rightarrow W^- \rightarrow ~ {\rm
anything}$.

This work, different from the previous ones (see, e.g., Refs.
\cite{Winter,XZ,BG,Serpico}) in studying how to determine or
constrain one or two of three neutrino mixing angles and the Dirac
CP-violating phase with neutrino telescopes, reveals the combined
effect of $\theta^{}_{13} \neq 0$, $\theta^{}_{23} \neq \pi/4$ and
$\delta \neq \pi/2$ which can show up at the detector. Even if
$\Delta \neq 0$ is established from the measurement of
ultrahigh-energy neutrino fluxes, the understanding of this
$\mu$-$\tau$ symmetry breaking signal requires more precise
information about $\theta^{}_{13}$, $\theta^{}_{23}$ and $\delta$.
Hence it makes sense to look at the complementary roles played by
neutrino telescopes and terrestrial neutrino oscillation
experiments (e.g., the reactor experiments to pin down the
smallest neutrino mixing angle $\theta^{}_{13}$ and the neutrino
factories or superbeam facilities to measure the CP-violating
phase $\delta$) in the era of precision measurements.

The feasibility of our idea depends on the assumption that we have
correctly understood the production mechanism of cosmic neutrinos
from a distant astrophysical source (i.e., via $pp$ and $p\gamma$
collisions) with little uncertainties. It is also dependent upon
the assumption that the error bars associated with the measurement
of relevant neutrino fluxes or their ratios are much smaller than
$\Delta$. The latter is certainly a challenge to the sensitivity
or precision of IceCube and other neutrino telescopes under
construction or under consideration, unless the effect of
$\mu$-$\tau$ symmetry breaking is unexpectedly large.
Nevertheless, there is no doubt that any constraint on $\Delta$ to
be obtained from neutrino telescopes will be greatly useful in
diagnosing the astrophysical sources and in understanding the
properties of neutrinos themselves. Much more efforts are
therefore needed to make in this direction.

\vspace{0.5cm}

The author would like to thank J.X. Lu for warm hospitality at the
Interdisciplinary Center for Theoretical Study of USTC, where part
of this paper was written. He is also grateful to H.B. Yu and S.
Zhou for useful discussions. This work is supported by the
National Natural Science Foundation of China.

\begin{figure}[t]
\vspace{-2.5cm}
\epsfig{file=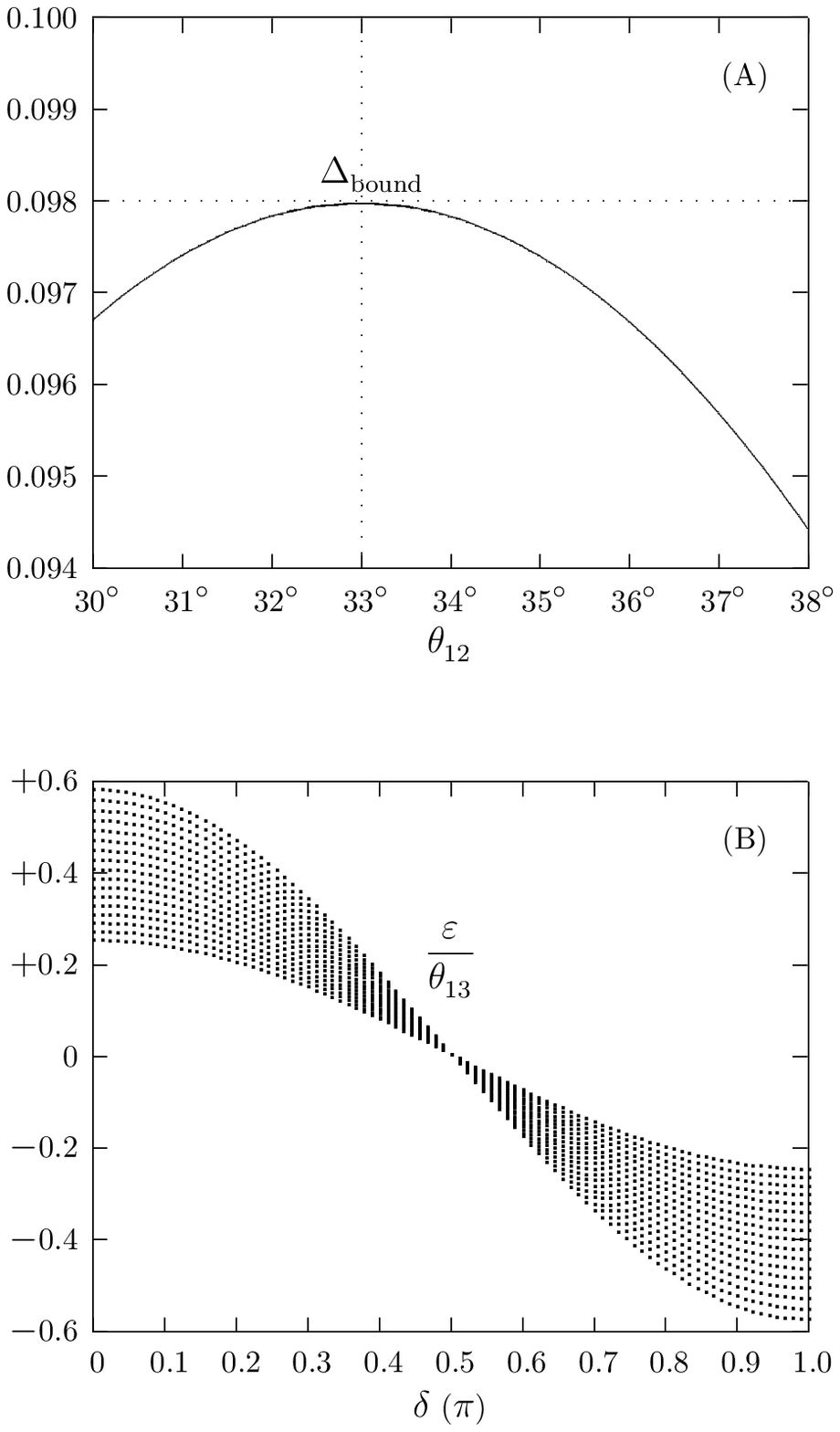,bbllx=1.5cm,bblly=17.5cm,bburx=11.5cm,bbury=28.5cm,%
width=9cm,height=10cm,angle=0,clip=0}
\vspace{8.6cm} \caption{(A)
the dependence of $\Delta^{}_{\rm bound}$ on $\theta^{}_{12}$,
where $\theta^{}_{13}$ and $\varepsilon$ take their respective
upper limits (i.e., $\theta^{}_{13} < 10^\circ$ and $\varepsilon <
9^\circ$); (B) the nontrivial condition for $\Delta =0$, where
$\theta^{}_{12}$ is allowed to vary in the range $30^\circ <
\theta^{}_{12} < 38^\circ$.}
\end{figure}

\end{document}